# Molecular mechanism of the Debye relaxation in monohydroxy alcohols revealed from rheo-dielectric spectroscopy

*Shalin Patil,[1] Ruikun Sun,[1] Shinian Cheng,[2] and Shiwang Cheng[1],\**

[1] Department of Chemical Engineering and Materials Science, Michigan State University, East Lansing, MI 48824, USA
[2] Institute of Physics, University of Silesia in Katowice, SMCEBI, 75 Pulku Piechoty 1A, 41-500 Chorzow, Poland

**Abstract**

Rheo-dielectric spectroscopy is employed, *for the first time*, to investigate the effect of external shear on Debye-like relaxation of a model monohydroxy alcohol, *i.e.* the 2-ethyl-1-hexanol (2E1H). Shear deformation leads to strong acceleration in the structural relaxation, the Debye relaxation, and the terminal relaxation of 2E1H. Moreover, the shear-induced reduction in structural relaxation time, $\tau_\alpha$, scales quadratically with that of Debye time, $\tau_D$, and the terminal flow time, $\tau_f$, suggesting a relationship of $\tau_D^2 \sim \tau_\alpha$. Further analyses reveal $\tau_D^2/\tau_\alpha$ of 2E1H follows Arrhenius temperature dependence that applies remarkably well to many other monohydroxy alcohols with different molecular sizes, architectures, and alcohol types. These results cannot be understood by the prevailing transient chain model and suggest a H-bonding breakage facilitated sub-supramolecular reorientation as the origin of Debye relaxation of monohydroxy alcohols, akin to the molecular mechanism for the terminal relaxation of unentangled "living" polymers.

\* Corresponding Author. Email Address: chengsh9@msu.edu

Monohydroxy alcohols, an important class of hydrogen-bonding (H-bonding) liquid, have an enormously strong electrical absorption at time scale longer than their structural relaxation for glass transition [1,2]. This intriguing strong electrical absorption leads to a distinct dielectric dispersion with a *single* relaxation time, *i.e.* the Debye relaxation, whose origin has been a subject of debate for more than 100 years [2-11]. Various models and theoretical understandings have been proposed to understand the emergence of the Debye-like relaxation in monohydroxy alcohols, including the transient chain model [7], dipole-dipole correlation [4,8,12], H-bonding associating dynamics [11], molecular rearrangement inside H-bonding clusters [3], density fluctuations [13], life-time of H-bonding [14,15], Maxwell-Wagner-Sillars' interfacial polarization [16], migration of defects [17], etc. However, none of them can adequately rationalize the two outstanding characteristics of Debye relaxation of monohydroxy alcohols: (i) the single-exponential dynamics that points to a non-collective feature and a lack of memory effect [18]; and (ii) the strong non-Arrhenius temperature dependence that represents the collective nature of the Debye relaxation [19].

The transient chain model [7], one of the most successful models by far for the Debye relaxation of monohydroxy alcohols, attributes the Debye relaxation to a supramolecular chain end-to-end vector relaxation. However, the supramolecular chain end-to-end vector relaxation must also generate a set of other sub-chain modes like the normal modes of type-A polymers [20], which have not been observed experimentally. To explain the single-relaxation time nature of the Debye relaxation, the transient chain model invokes additional mechanism of the alcohol association/dissociation at the supramolecular chain ends [7]. However, it remains unclear how the alcohol association/dissociation via H-bonding connects to the single-relaxation time Debye process since the H-bonding lifetime is demonstrated to be different from the Debye relaxation time [6]. Therefore, fundamental questions remain open on the relationship between the reversible H-bonding breakup and reformation, the supramolecular structures formation, and the Debye relaxation.

In this Letter, we employ rheo-dielectric spectroscopy, *for the first time*, to investigate the molecular origin of Debye relaxation of monohydroxy alcohols, targeting a *quantitative* understanding of the relationship between the reversible H-bonding breakup and reformation and the Debye relaxation. Different from *almost all* previous dielectric measurements, we apply external shear deformation to perturb the liquid structures for insights on the Debye relaxation. A model monohydroxy alcohol, *i.e.,* the 2-ethyl-1-hexanol (2E1H), was chosen in the study due to its well-resolved Debye and structural relaxation. In addition, a polymer, poly(propylene glycol) of molecular weight of 4 kg/mol (PPG4K), is included due to its comparable separation between the structural relaxation time, $\tau_\alpha$, and the terminal relaxation time, $\tau_f$, with 2E1H. **Figure 1a** presents the experimental protocol of combining rheology and dielectric spectroscopy, where the rheological response and dielectric properties were collected simultaneously right after oscillatory shear (OS) of a strain amplitude of $\gamma_0$ and an angular frequency of $\omega$. Detailed description of the rheo-dielectric spectroscopy is described in the **Supplementary Materials (SM)**. **Figure 1b** presents the time-dependent storage modulus, $G'(t)$, during the OS step at $\gamma = 10\ \%$ and $\omega = 10\ rad/s$ up to time $t_1 = 1{,}000\ s$ for both PPG4K and 2E1H. In these experiments, we match $\tau_f$ of 2E1H and PPG4K by choosing different testing temperatures of $T = 170\ K$ for 2E1H and $T = 220\ K$ for PPG4K. While there are almost no changes in $G'(t)$ of PPG4K, a noticeable drop of $G'(t)$ is observed for 2E1H in the OS step at the long-time limit, highlighting an interesting influence of shear to the structure and dynamics of 2E1H.

At $T = 170\ K$, $\tau_\alpha \approx 2 \times 10^{-4}\ s$ for 2E1H. At $T = 220\ K$, $\tau_\alpha \approx 1.6 \times 10^{-4}\ s$ for PPG4K. The corresponding structural relaxation time Weissenberg numbers are $Wi_\alpha = \dot\gamma \tau_\alpha \approx 2 \times 10^{-3} \ll 1$ for 2E1H and $Wi_\alpha = \dot\gamma \tau_\alpha \approx 1.6 \times 10^{-3} \ll 1$ for PPG4K at shear rate $\dot\gamma = 10\ s^{-1}$. Had no shear-induced structural modifications, one should not anticipate any changes in the structural relaxation [21-23], as shown in **Figures 1c** and **1d** of the comparison of dielectric loss spectra, $\varepsilon''(\omega)$, and rheological spectra, $G'(\omega)$ and $G''(\omega)$, of PPG4K before ($G'(\omega)$: red circles, and $G''(\omega)$: grey circles) and after ($G'(\omega)$: blue squares) OS. Moreover, the dielectric normal modes and the terminal relaxation of PPG4K are not affected by shear, consistent with the previous measurements of rheo-dielectric measurements of polymers [24]. In contrast, significant shift of the structural relaxation to higher frequencies of 2E1H have been observed in $\varepsilon''(\omega)$. At the same time, the terminal relaxation of 2E1H from rheology and Debye relaxation from dielectric measurement shift to higher frequencies, as shown in **Figures 1e** and **1f.**

To quantify the shear-induced changes, we fit dielectric spectra of both 2E1H and PPG4K with Havriliak-Negami functions (the dashed lines in **Figures 1c-1f**). Several features are worth noting: (i) External shear influences *little* the shapes of both Debye relaxation and the structural relaxation; (ii) An around four-time reduction in $\tau_\alpha$ and an around two-time reduction in $\tau_D$ are observed at $\gamma = 10\ \%$ and $\omega = 10\ rad/s$, highlighting the different responses of structural and Debye relaxations to external deformation and an interesting shear-induced *further separation* between $\tau_D$ and $\tau_\alpha$. The shear-induced further separation of $\tau_D$ and $\tau_\alpha$ can be better demonstrated through the normalized dielectric spectra (**Figure S2** in **SM**). (iii) The dielectric amplitude of structural relaxation, $\Delta\varepsilon_\alpha$, remain *little* changed with shear (from $\Delta\varepsilon_\alpha \sim 0.52$ before deformation to $\Delta\varepsilon_\alpha \sim 0.54$ after deformation), while a clear increase is observed in the dielectric amplitude of Debye relaxation (from $\Delta\varepsilon_D = 24.4$ before deformation to $\Delta\varepsilon_D = 26.4$ after deformation) (**Figure**

**S3 in SM**). (iv) The rheological flow time, $\tau_f$, of 2E1H moves in accord with the Debye relaxation time, $\tau_D$, with shear emphasizing their similar responses to shear deformation. The observed identical shear dependence of $\tau_f$ and $\tau_D$ supports a previous viewpoint about a strong correlation between $\tau_f$ and $\tau_D$ [5,25], as discussed further later.

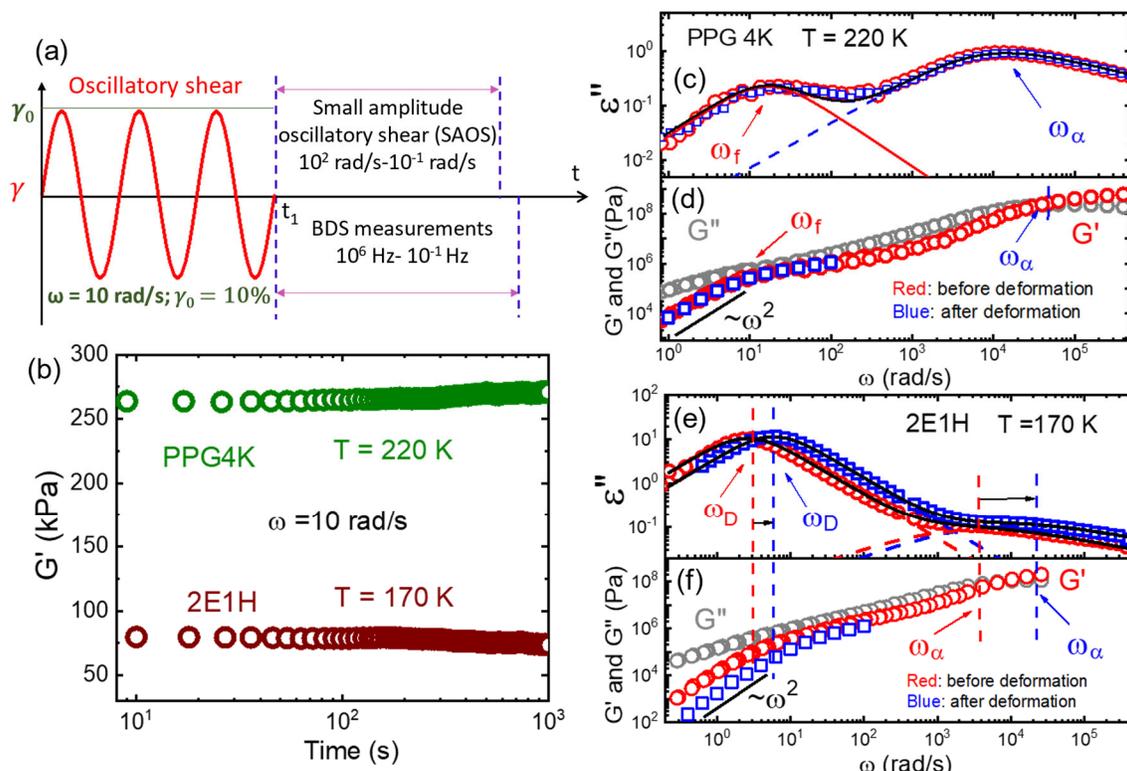

**Figure 1.** (a) A sketch of the experimental protocol with pre-deformation of oscillatory shear (OS) followed by rheology or dielectric measurements. (b) Time-dependent storage modulus, $G'(\omega;t)$, of 2E1H and PPG4K at the OS step. (c) Dielectric loss spectra of PPG4K before (red circle) and after (blue squares) OS. (d) Rheology spectra of PPG4K before (circles) and after (squares) OS. (e) Dielectric loss spectra of 2E1H before (red circle) and after (blue squares) OS. (f) Rheology spectra of 2E1H before (circles) and after (squares) OS.

**Figure 1** represents the first set of our key findings with many features not being able to be understood by existent theories and models. H-bonding interactions of monohydroxy alcohols can induce supramolecular chain formation. The acceleration in $\tau_\alpha$ and $\tau_D$ might be explained through shear-induced supramolecular chain destruction per transient chain model. However, according to the transient chain model, the supramolecular chain destruction should lead to a shortening in supramolecular chain length and a decrease in $\tau_D/\tau_\alpha$ or $\Delta\varepsilon_D/\Delta\varepsilon_\alpha$ [5,7] that is at odds with the experimental observation. From this perspective, it is challenging to anticipate the acceleration in $\tau_\alpha$ and $\tau_D$, and *simultaneously* an increment in $\tau_D/\tau_\alpha$ or $\Delta\varepsilon_D/\Delta\varepsilon_\alpha$.

To explore the shear conditions to the Debye relaxation and structural relaxation of monohydroxy alcohols, we vary shear rates, $\dot{\gamma} = 0.1 - 10 \, s^{-1}$, while fixing the applied strain amplitude $\gamma = 10\%$ and testing temperature at $T = 170 \, K$. The time of OS step is also fixed at $t_1 = 1,000 \, s$. The upper inset panel of **Figure 2** summaries the corresponding Debye time and structural relaxation at different shear rates, $\tau_D(\dot{\gamma})$ and $\tau_\alpha(\dot{\gamma})$, where higher applied shear rates tend to have smaller values of $\tau_D(\dot{\gamma})$ and $\tau_\alpha(\dot{\gamma})$. Moreover, the shear induces shifts in Debye relaxation time, $\tau_D(\dot{\gamma})/\tau_D(\dot{\gamma} = 0)$, and the structural relaxation time, $\tau_\alpha(\dot{\gamma})/\tau_\alpha(\dot{\gamma} = 0)$, follow an interesting relationship of $\tau_D(\dot{\gamma})/\tau_D(\dot{\gamma} = 0) \sim (\tau_\alpha(\dot{\gamma})/\tau_\alpha(\dot{\gamma} = 0))^{1/2}$ as demonstrated in the main frame of **Figure 2**. This implies a strong connection between $\tau_D^2$ and $\tau_\alpha$ with the ratio of $\tau_D^2/\tau_\alpha$ being invariant with shear (bottom inset panel of **Figure 2**). We emphasize that the revealed intrinsic relation between $\tau_D^2$ and $\tau_\alpha$ is fundamentally different from that of the transient chain model [7] and the recently proposed dipole-dipole cross correlation [4,8,26], both of which connect $\tau_D$ and $\tau_\alpha$ through $\tau_D/\tau_\alpha$.

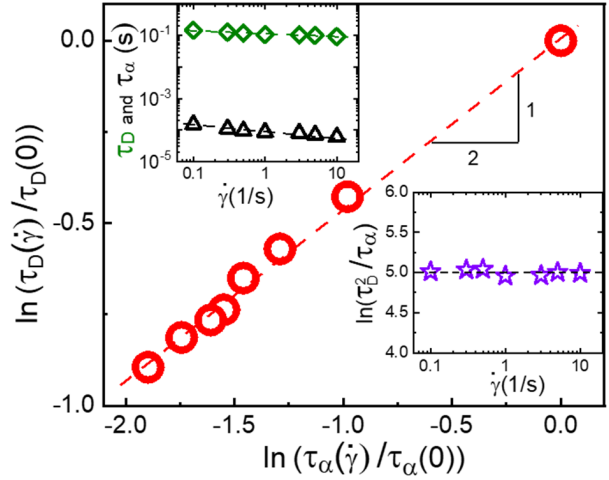

**Figure 2.** The relationship of the shift in Debye relaxation and structural relaxation with shear. The upper inset panel shows the values of $\tau_D$ and $\tau_\alpha$. The upper inset shows the shifts of $\tau_D$ and $\tau_\alpha$ with shear. The bottom inset panel shows the $ln(\tau_D^2/\tau_\alpha)$ vs $\dot{\gamma}$. The dashed lines are guides for eyes. The dashed black line in the bottom inset represent the $ln(\tau_D^2/\tau_\alpha)$ values in the absence of shear.

To understand the revealed relationship between $\tau_D^2$ and $\tau_\alpha$, we focus on the reversible H-bonding dynamics and supramolecular structures of 2E1H. Specifically, the reversible H-bonding breakup and reformation reminds of a type of "living" polymers [27], whose monomers can freely associate/dissociate from the polymer. According to Cates and co-workers [28-30], the reversibility of monomer bonding leads to fundamentally different dynamics of "living" polymers from conventional covalently bonded polymers (for more details see **SM**). In a fast chain breakage limit of $\tau_B \ll \tau_c$, a "living" polymer would break and reform many times before its end-to-end vector relaxation time $\tau_c$, where $\tau_B = \frac{1}{k_2 \bar{N}}$ is the average chain breakage time with $k_2$ being the reaction rate constant of monomer dissociation and $\bar{N}$ the characteristic chain length of the living polymer. Note that the chain break time $\tau_B$ is $\bar{N}$ times shorter than the lifetime of a reversible bond since breakage of any one of the $\bar{N}$ reversible bonds will lead to the breakage of the chain. As a result, the terminal relaxation (or flow time) of an entangled "living" polymer, $\tau_f$, can be much shorter than its end-to-end reorientation time, $\tau_c$, with $\tau_f = (\tau_B \tau_c)^{1/2}$ that agrees widely with experiments [27,28]. More specifically, the terminal relaxation of "living" polymers is dictated by orientational dynamics of a *sub-chain* rather than the end-to-end vector relaxation of the *whole* chain. The fast chain breakage also leads to an exceptional narrow *single relaxation mode* terminal relaxation process [27,29] (also see **Section 6.1** of the **SM** for discussions on the physical origin of the emergence of the *single relaxation time mode*).

Cates' living polymer model was constructed for entangled polymers. However, the rheological measurements **(Figure 1f)** clearly suggest the absence of entanglement in 2E1H and many other monohydroxy alcohols[31]. We thus extend Cates' living polymer analysis to unentangled systems (see **Section 6.2** of **SM**) and obtain the terminal relaxation time of unentangled "living" polymer as $\tau_f = (\tau_c \tau_B)^{1/2}$ in the fast chain breakage limit. Here, $\tau_c = \tau_\alpha \bar{N}^2$ is the longest end-to-end vector relaxation time of the an unentangled supramolecular chain, $\bar{N} = \sqrt{\frac{c_0 k_1}{2k_2}}$ is the characteristic supramolecular chain length, and $c_0$, $k_1$, and $k_2$ are the molar concentration of alcohol molecules in the supramolecular chains, the reaction rate constant of the H-bonding association, and the rate constant of H-bonding dissociation. We note that the living polymer analysis emphasizes a sub-chain reorientation *rather than* the whole chain end-to-end vector orientation responsible for the terminal relaxation of the polymer that should be a single relaxation time process. Given the polarity of H-bonding, dipole accumulates along the backbone of supramolecular chains of monohydroxy alcohols. The near single relaxation time *sub-chain* reorientation process should thus lead to an active Debye-like dielectric process on the time scale of $\tau_f$. Experimentally, $\tau_D \approx \tau_f$ has been observed for many monohydroxy alcohols [5,25] and also see **Figures 1e-1f**. Therefore, the above analyses unravel a molecular mechanism for the Debye relaxation of monohydroxy alcohols and provide a theoretical justification of the widely observed $\tau_D \approx \tau_f$ of monohydroxy alcohols.

Can the analogy to "living" polymer explain the scaling of $\tau_D^2 \sim \tau_\alpha$ revealed by rheo-dielectric spectroscopy? For unentangled polymers, the end-to-end vector relaxation time $\tau_c = \tau_\alpha \bar{N}^2$. Combining $\tau_f \approx \tau_D$ and $\tau_f = (\tau_c \tau_B)^{1/2}$, one has:

$$\tau_D^2 = \tau_c \tau_B = \tau_\alpha \bar{N}^2 \times \frac{1}{k_2 \bar{N}} = \frac{\tau_\alpha \bar{N}}{k_2} = \tau_\alpha \sqrt{\frac{c_0 k_1}{2 k_2^3}} \quad (1)$$

Rewritting Eqn.1, one obtains

$$\frac{\tau_D^2}{\tau_\alpha} = \sqrt{\frac{c_0 k_1}{2 k_2^3}} \quad (2)$$

Since $k_1$, and $k_2$ are reaction rate constant and mostly sensitive primarily to testing temperature, Eqn.2 predicts $\tau_D^2/\tau_\alpha$ an invariant with shear, consistent with experimental observations (inset of **Figure 2**). Thus, the analogy to "living" polymer explains the intriguing scaling of $\tau_D^2 \sim \tau_\alpha$ and the shear induced acceleration in alcohol dynamics.

Furthermore, assuming Arrhenius laws of $k_1 \sim \exp(-\frac{E_1}{RT})$ and $k_2 \sim \exp(-\frac{E_2}{RT})$ with $E_1$ and $E_2$ being the activation energies and $R$ the gas constant, Eqn.2 implies

$$\tau_D^2/\tau_\alpha \sim \exp(-(E_1 - 3E_2)/(2RT)) \quad (3)$$

that gives an Arrhenius temperature dependence of $\tau_D^2/\tau_\alpha$ with an apparent activation energy, $\Delta E = (3E_2 - E_1)/2$. Note that Eqn. 3 is a prediction of our extension of "living" polymer model to unentangled monohydroxy alcohols, which does not depend on shear and should apply to other monohydroxy alcohols forming unentangled supramolecular chains. Since $\tau_D$ and $\tau_\alpha$ can be accurately identified through dielectric measurements over a wide temperature range, Eqn. 3 can also be tested experimentally that also provides an examination of the proposed molecular understanding.

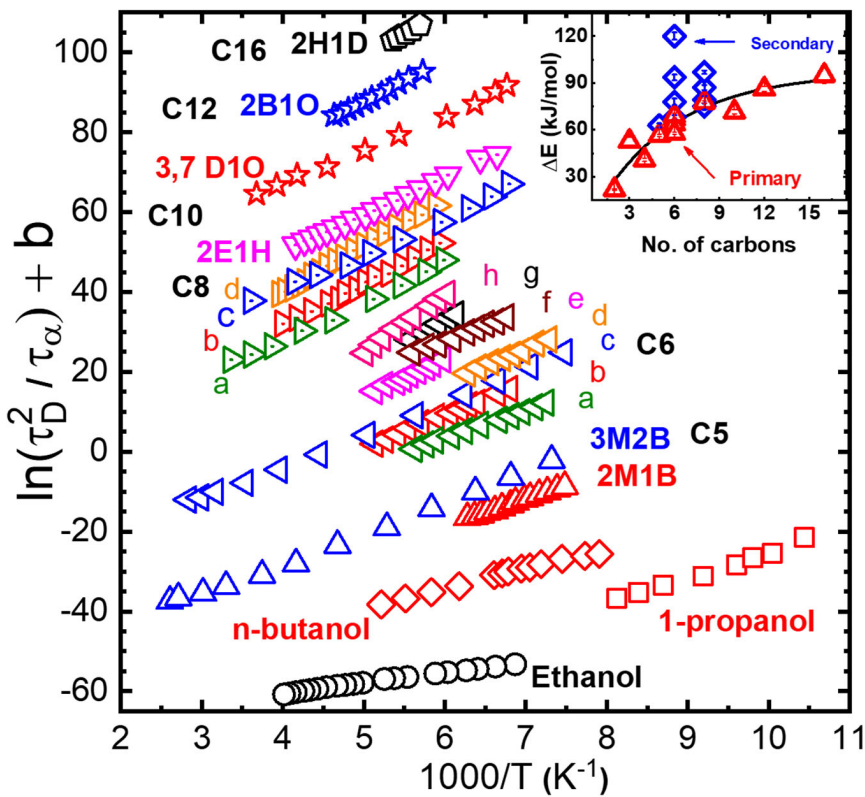

**Figure 3**. Plots of $ln(\tau_D^2/\tau_\alpha)$ vs $1000/T$ for 2E1H and other monohydroxy alcohols. The Y-axis is shift by an amplitude of $b$ for the presentation purposes. The details of $b$ values are given in **Table S1** of the supplementary materials (SM). These monohydroxy alcohols are: EL: ethanol [16]; 1-PL: 1-propanol [16]; 1-BL: n-butanol [16]; 2M1B: 2-methyl-1-butanol [16]; 3M2B: 3-methyl-2-butanol [32], C6a: 3-methyl-1-pentanol (3M1P) [33]; C6b: 2-hexanol [33]; C6c: 2-ethyl-1-butanol (2E1B) [33]; C6d: 4-methyl-1-pentanol [33]; C6e: 3-methyl-2-pentanol (3M2P) [33]; C6f:2-methyl-1-pentanol (2M1P) [33]; C6g: 4-methyl-2-pentanol (4M2P) [33]; C6h: 2-methyl-3-pentanol (2M3P) [33]; C8a: 4-methyl-3-heptanol (4M3H) [34]; C8b:5-methyl-3-heptanol (5M3H) [34]; C8c: 5-methyl-2-heptanol (5M2H) [34]; C8d: 6-methyl-3-heptanol (6M3H) [34]; 2E1H: 2-ethyl-1-hexanol [25]; 3,7D1O: 3,7-Dimethyl-1-octanol (3,7D1O) [32]; 2B1O: 2-butyl-1-octanol (2B1O) [35]; 2H1D: 2-hexyl-1-dodecanol [35]. The label $Cx$ with $x = 5, 6, 8, 10, 12,$ and 16 referring to the number of carbon atom of the alcohol molecules. Inset shows the apparent activation energy, $\Delta E$, of each alcohol estimated from the Arrhenius plots.

**Figure 3** plots the temperature dependence of $\tau_D^2/\tau_\alpha$ of 2E1H (the pink dot down triangles) and many other monohydroxy alcohols. The Y-axis of **Figure 3** is shifted vertically by an amplitude of $b$ for presentation purposes (the $b$ values are presented in **Table S1** of **SM**). An Arrhenius plot of $\tau_D^2/\tau_\alpha$ is indeed obtained over an *exceptional* wide temperature range ($T = 240 - 150\ K$) and time scales ($\tau_\alpha \approx 2.1 \times 10^{-9} - 8 \times 10^{-1}$) for 2E1H, supporting the prediction of Eqn 3. Remarkably, similar Arrhenius temperature dependence of $\tau_D^2/\tau_\alpha$ can be obtained for many other monohydroxy alcohols of different molecular sizes (the number of carbon molecules in the backbone, from ethanol (C2) to 2-hexyl-1-dodecanol (C16)), the types of alcohols (primary or secondary), and the molecular architectures (linear or branches). The broad agreement between experiments and Eqn.3 highlights the *universality* of the Arrhenius temperature dependence of $\tau_D^2/\tau_\alpha$ and the predictive power of the new understanding of Debye relaxation. Furthermore, previous studies based on transient chain model suggest $\tau_D/\tau_\alpha$ or $\Delta\varepsilon_D/\Delta\varepsilon_\alpha$ connecting to the supramolecular chain size [6,7], which should increase with cooling given the exothermal H-bonding formation reaction [36,37]. However, $\tau_D/\tau_\alpha$ exhibits non-monotonic temperature dependence and $\Delta\varepsilon_D/\Delta\varepsilon_\alpha$ reaches a saturation approaching to the glass transition temperature $T_g$ [32] in experiments (see **Figure S4** in **SM**), both of which are at odds with the transient chain model. One the other hand, the revealed Arrhenius temperature dependence of $\tau_D^2/\tau_\alpha$ holds across an exceptional wide temperature range and time scales, resolving the relationship between $\tau_D$ and $\tau_\alpha$ and their connection with the supramolecular structures.

Furthermore, apparent activation energies, $\Delta E$, of monohydroxy alcohols can be obtained from the Arrhenius plots. The inset of **Figure 3** presents a summary of $\Delta E$ with $\Delta E \approx 25 - 120 \, kJ/mol$ varying with the alcohol sizes, alcohol types, and architectures. As expected, these apparent activation energies, $\Delta E = (3E_2 - E_1)/2 = E_2 + \Delta H/2 > \Delta H/2$ are all higher than half of the enthalpy of H-bonding formation with $\Delta H/2 \approx (E_2 - E_1)/2 \approx 5 - 10 \, kJ/mol$ [38]. Another interesting observation is that $\Delta E$ of primary alcohols seems to increase with alcohol sizes with a saturation at alcohol sizes beyond 8-10 carbon atoms in the backbone. Since $E_2$ depends on the structures of alcohols, it is not surprising to see the chemistry dependence of $\Delta E$. More experiments are needed to confirm the observations for future investigations.

The above analyses demonstrate the relevance of the "living" polymer analogy to understand the supramolecular dynamics and Debye relaxation of monohydroxy alcohols. The rheo-dielectric measurements reveal two other interesting features—a reduction in $\tau_\alpha$ and an increment in $\Delta\varepsilon_D$ with shear. Note that the "living" polymer model says nothing about the glassy dynamics and the dielectric amplitude. Are the observed changes in $\tau_\alpha$ and $\Delta\varepsilon_D$ compatible with the proposed molecular understanding of Debye relaxation? According to the dielectric theory [39,40], $\Delta\varepsilon_D \approx Fg\frac{N}{V}\frac{C_\infty \mu_m^2}{3\varepsilon_0 k_B T}$ with $F$ being the Onsager factor, $g$ the Kirkwood-Fröhlich factor, $\varepsilon_0$ the vacuum permittivity, $N/V$ the number density of the alcohol molecules in supramolecular chains, $C_\infty$ the characteristic ratio of the supramolecular chain, and $\mu_m$ the individual alcohol dipolar moment along the chain backbone. The analyses indicate that supramolecular chain length does not directly affect $\Delta\varepsilon_D$ and $\Delta\varepsilon_D/\Delta\varepsilon_\alpha$ should not be a good indicator of supramolecular chain length. Since little changes in $C_\infty$ and $\mu_m$ should be anticipated under the mild shear conditions of $\dot{\gamma}\tau_D \sim 1$, the most probable mechanism leading to the noticeable increment in $\Delta\varepsilon_D$ is a shear-induced increment in $N/V$. We note that H-bonding can induce both supramolecular chain formation and supramolecular ring formation, and the supramolecular rings do not have active dielectric dispersion. Thus, it is possible that the applied shear induces ring-to-chain transition of monohydroxy alcohols resulting in an increment of $N/V$. Since ring polymers should have much higher $T_g$ than the linear counterparts in the unentangled region [34,41], the shear-induced ring-to-chain transition should also lead to an reduction in $T_g$ and an acceleration in $\tau_\alpha$, both of which agree with the observations. We emphasize that external perturbation induced ring-to-chain transition has been observed previously in monohydroxy alcohols [10].

In summary, rheo-dielectric spectroscopy has been performed to elucidate the supramolecular dynamics and Debye relaxation of monohydroxy alcohols. Dielectric measurements of 2E1H reveal strong shear-induced acceleration in the structural relaxation time, $\tau_\alpha$, Debye relaxation time, $\tau_D$, and an increase in the dielectric amplitude of Debye process, $\Delta\varepsilon_D$. At the same time, rheological measurements exhibit speeding up in terminal relaxation and a strong coupling between $\tau_f$ and $\tau_D$ with $\tau_D \approx \tau_f$. The shear-induced shifts in $\tau_\alpha$ scales quadratically with that of $\tau_f$ or $\tau_D$. Detailed analyses reveal Arrhenius temperature dependence of $\tau_D^2/\tau_\alpha$ of 2E1H in the absence of shear that applies remarkably well for a large number of monohydroxy alcohols with different molecular sizes, alcohol types, and molecular architectures. These features cannot be explained by the prevailing transient chain model and the recently proposed dipole-dipole cross correlation mechanism. We rationalize them through *extending* Cates' living polymer analysis to unentangled living polymers that suggest the reorientation dynamics of sub-chains rather than the end-to-end vector relaxation of the whole supramolecular chain as the origin of Debye relaxation of monohydroxy alcohols. The combined experiments and modeling thus provide a new perspective for the dynamics of monohydroxy alcohols, and point to a paradigm shift for the understanding of Debye relaxation as well as its relationship with the reversible dynamics of H-bonding breakup and reformation of monohydroxy alcohols.

**Acknowledgement**

This work was supported by Michigan State University (MSU) Discretionary Funding Initiative. We thank the discussions with Prof. Ron Larson on the dynamics of unentangled living polymers.

SUPPLEMENTARY MATERIALS

# Molecular mechanism of the Debye relaxation in monohydroxy alcohols revealed from Rheo-dielectric spectroscopy

*Shalin Patil, Ruikun Sun, Shinian Cheng, and Shiwang Cheng*
Author Correspondence should be addressed to Shiwang Cheng at <chengsh9@msu.edu>

## 1. Materials and Methods.

**1.1 Materials.** Two materials are included in the experiments, 2-ethyl-1-hexanol (2E1H, Sigma-Aldrich, ≥ 99.6%), and poly(propylene glycol) with a number average molecular weight of 4 kg/mol (PPG4K, Sigma-Aldrich). Both samples were used as received.

**1.2 Rheo-dielectric spectroscopy.** The rheo-dielectric spectroscopy is constructed through a combination of an Advanced Rheometric Expansion System (ARES, TA instrument) rotational rheometer and a Novocontrol Alpha analyzer with a ZG4 testing interface. The ARES rheometer equips with a Rheometric Scientific Oven for temperature control with an accuracy of $\pm 0.1$ K. The Novocontrol Alpha analyzer with a ZG4 extension test interface covers dielectric dispersion measurement across a frequency range of $10^6$ Hz to $10^{-2}$ Hz. The rheo-dielectric measurements were performed on a pair of home-made parallel plates of diameter of 8 mm that are electronically insulated from the ARES rheometer. A gap of ~0.4 mm is applied in all measurements. In rheo-dielectric measurements, the rheometer and the dielectric measurements can be programmed to capture simultaneously the rheological response and dielectric properties during and after deformation. In the current study, we follow the protocols discussed in **Figure 1a** of the main context for rheo-dielectric measurements. The measurements for 2E1H were at $T = 170\ K$ and for PPG at $T = 220\ K$ to match their terminal relaxation time, $\tau_f$.

The dielectric spectra of 2E1H were analyzed through a combination of a Debye function and a Havriliak- Negami (HN) function:

$$\varepsilon^*(\omega) = \varepsilon_\infty + \frac{\Delta\varepsilon_D}{1 + i\omega\tau_{HN,D}} + \frac{\Delta\varepsilon_\alpha}{\left(\left(1 + i\omega\tau_{HN,\alpha}\right)^{\beta_\alpha}\right)^{\gamma_\alpha}} + \frac{\sigma_{DC}}{i\omega\varepsilon_0} + A\omega^{-n}$$

where $i$ is the imaginary unit, $\varepsilon^*$ is the complex permittivity, $\varepsilon_\infty$ and $\varepsilon_0$ are the dielectric constant at an infinite high frequency and the vacuum permittivity, $\omega$ is the angular frequency, $\tau_{HN,D}$ and $\tau_{HN,\alpha}$ are the characteristic HN time of the Debye relaxation and the structural relaxation, $\Delta\varepsilon_D$ and $\Delta\varepsilon_\alpha$ are the dielectric amplitudes of the Debye process and the structural relaxation process, $\sigma_{DC}$ is the dc-conductivity, $\beta_\alpha$ and $\gamma_\alpha$ are the shape parameters of the structural relaxation, and $A$ and $n$ are fit constants.

The dielectric spectra of PPG4K were analyzed through two HN functions: one for the normal mode (N) and the other for the structural relaxation of PPG4K:

$$\varepsilon^*(\omega) = \varepsilon_\infty + \frac{\Delta\varepsilon_N}{\left((1+i\omega\tau_{HN,N})^{\beta_N}\right)^{\gamma_N}} + \frac{\Delta\varepsilon_\alpha}{\left((1+i\omega\tau_{HN,\alpha})^{\beta_\alpha}\right)^{\gamma_\alpha}} + \frac{\sigma_{DC}}{i\omega\varepsilon_0} + A\omega^{-n}$$

The characteristic relaxation time of the $k^{th}$ process ($k = D, N, \text{and } \alpha$) can be obtained from the characteristic HN time of the structural relaxation:

$$\tau_k = \tau_{HN,k}\left[\sin\frac{\beta_k\pi}{2+2\gamma_k}\right]^{-1/\beta_k}\left[\sin\frac{\beta_k\gamma_k\pi}{2+2\gamma_k}\right]^{1/\beta_k}$$

For Debye process, $\tau_D = \tau_{HN,D}$.

The structural relaxation time, $\tau_\alpha$, and terminal relaxation time (or flow time), $\tau_f$, of 2E1H and PPG4K can also be determined from linear viscoelastic master curves constructed through applying the time-temperature superposition principle. The $\tau_\alpha$ is obtained through the high-frequency crossover, $\omega_\alpha$, between the storage modulus, $G'(\omega)$, and loss modulus, $G''(\omega)$. The terminal relaxation time, $\tau_f$, is obtained through following a previous method [1] from the onset frequency of shear thinning, $\omega_f$, with $\tau_f = 1/\omega_f$ that is also the highest frequency following the $G'(\omega) \sim \omega^2$. **Figure S1** presents demonstrations of the identification of $\tau_\alpha$ and $\tau_f$ of 2E1H and PPG4K from rheology. In this study, we used the $\tau_\alpha$ of BDS measurements since both $\tau_\alpha$ and $\tau_D$ can be identified directly from BDS measurements.

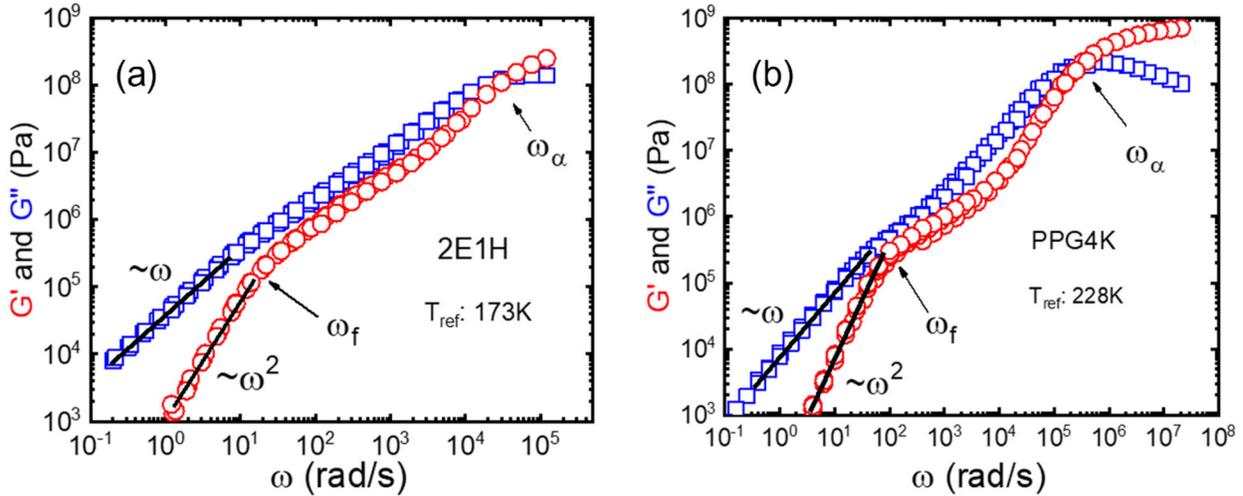

**Figure S1.** Identification of the structural relaxation time, $\tau_\alpha$, and the terminal relaxation, $\tau_f$, of (a) 2E1H and (b) PPG4K from rheological measurements. The master curves of 2E1H and PPG4K were constructed by following the time-temperature superposition principle. The $\tau_f$ is defined through the angular frequency, $\omega_f$, at which the terminal region starts to emerge with $G' \sim \omega^2$. A similar method have been proposed in [1] for identifying the terminal relaxation time of monohydroxy alcohols.

**2. Comparison of normalized spectra of 2E1H before and after shear**

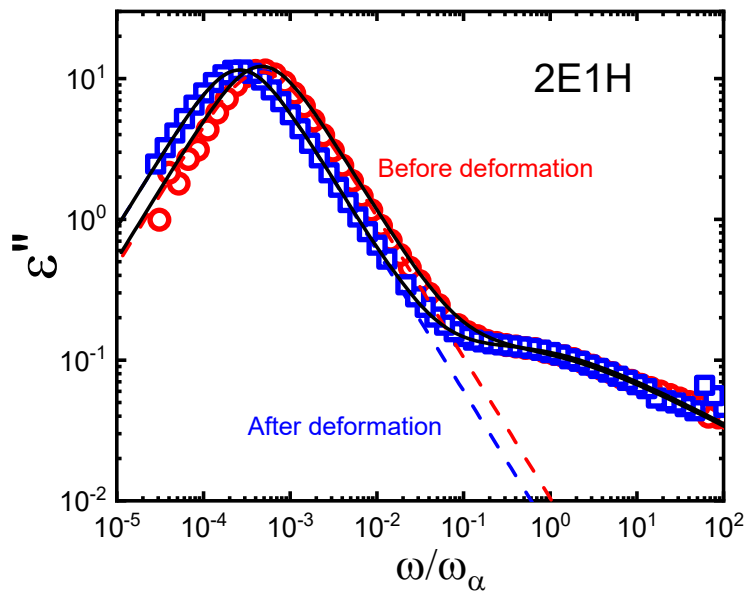

**Figure S2.** Normalized dielectric loss spectra, $\varepsilon''(\omega)$, of 2E1H before and after shear at $T = 170\ K$ with respect to $\omega/\omega_\alpha$. The shapes of the structural relaxation remain unchanged with shear, and a clear shear-induced separation between Debye relaxation and structural relaxation can be observed. The dashed lines represent fits to the Debye processes.

## 3. Dielectric storage spectra, $\varepsilon'(\omega)$, of 2E1H before and after shear.

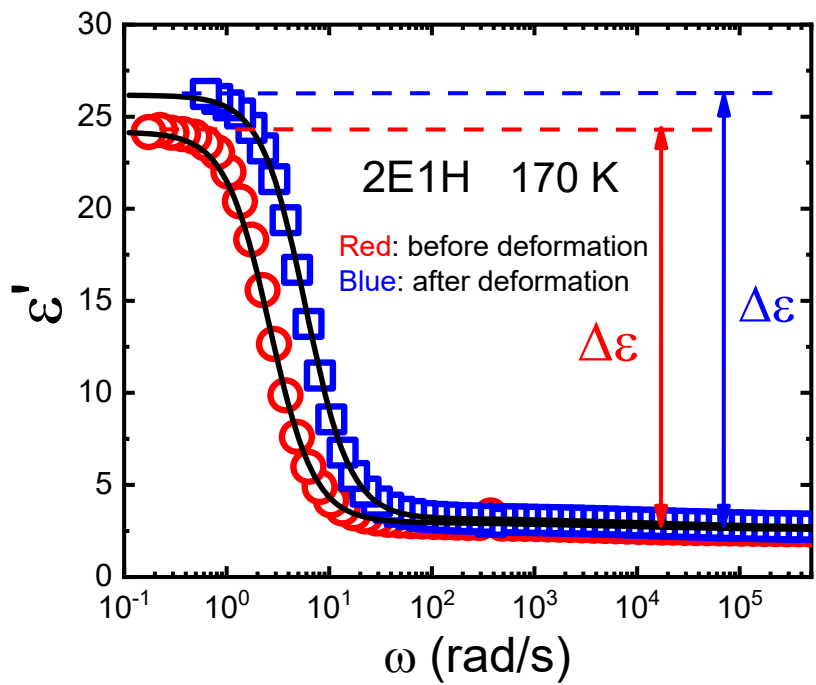

**Figure S3.** Dielectric storage spectra, $\varepsilon'(\omega)$, of 2E1H before and after oscillatory shear at strain amplitude $\gamma_0 = 10\ \%$ and angular frequency $\omega = 10\ rad/s$. A clear increment in the dielectric

amplitude of the Debye process can be identified. The black lines represent fits using the HN functions.

## 4. Variation of $\tau_\alpha$ and $\tau_D$ as a function of temperature

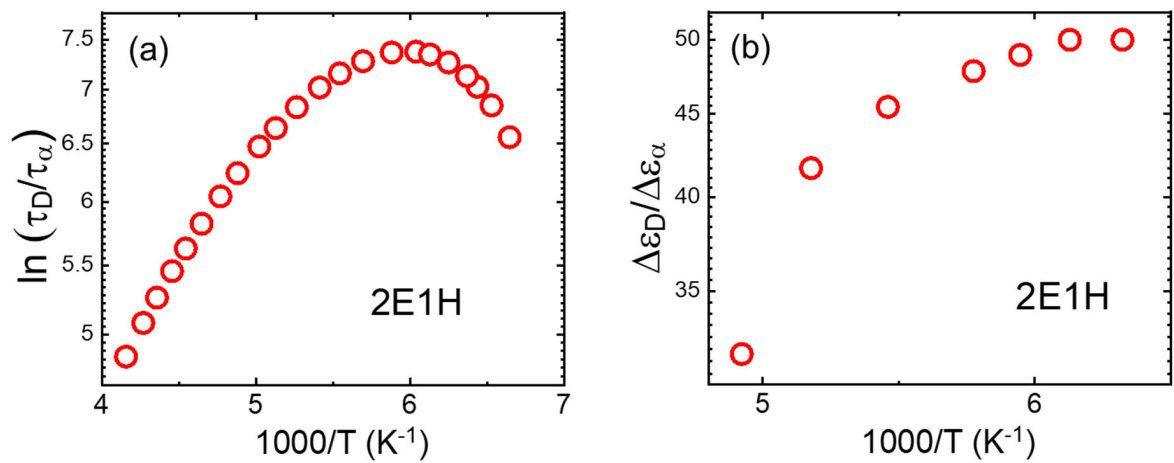

**Figure S4.(a)** Non-monotonic temperature dependence of logarithmic ratio of $\tau_D$ and $\tau_\alpha$ and ratio of dielectric amplitude of the Debye and the alpha process. (b) The ratio of dielectric amplitude of the Debye process and the structural relaxation process, $\Delta\varepsilon_D/\Delta\varepsilon_\alpha$, at different temperatures.

## 5. Molecular structures and abbreviations of monohydroxy alcohols.

Table S1. Molecular Structures and Abbreviation of various monohydroxy alcohols

| Abbreviation* | Alcohol name | Molecular Structure | Total carbon | Apparent Activation energy (KJ/mol) | Shift factor b† | Ref |
|---|---|---|---|---|---|---|
| EL (P) | Ethanol | | 2 | 21.6 | -45 | [2] |
| 1-PL (P) | Propanol | | 3 | 52.6 | -35 | [2] |
| 1-BL (P) | n-butanol | | 4 | 40.7 | -28 | [2] |
| 2M1B (P) | 2-methyl-1-butanol | | 5 | 56.4 | -17 | [2] |
| 3M2B (S) | 3-methyl-2-butanol | | 5 | 62.8 | -13 | [3] |
| 3M1P (P) | 3-methyl-1-propanol | 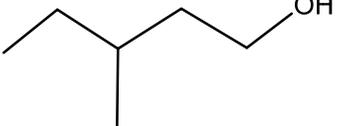 | 6 | 57.7 | +6 | [4] |
| 2E1B (P) | 2-ethyl-1-butanol | 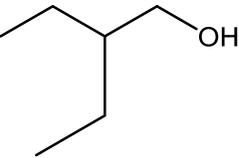 | 6 | 67.1 | +8 | [4] |
| 4M1P (P) | 4-methyl-1-pentanol | 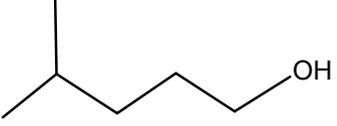 | 6 | 68.8 | +22 | [4] |
| 3M2P (S) | 3-methyl-2-pentanol | 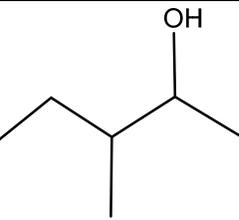 | 6 | 77.9 | +16 | [4] |
| 2M1P (P) | 2-methyl-1-pentanol | 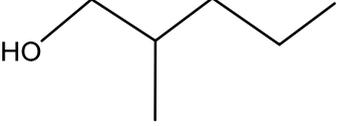 | 6 | 63.8 | +28 | [4] |
| 4M2P (S) | 4-methyl-2-pentanol | 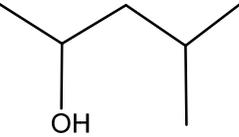 | 6 | 93.5 | +30 | [4] |
| 2M3P (S) | 2-methyl-3-pentanol | 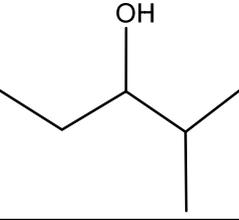 | 6 | 119.9 | +28 | [4] |
| 2-HL (P) | 2-hexanol | 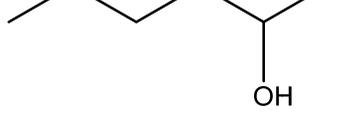 | 6 | 68.0 | +8 | [4] |
| 4M3H (S) | 3-methyl-3-heptanol | 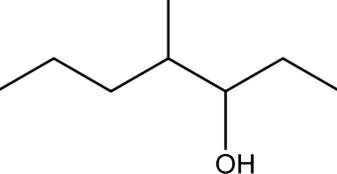 | 8 | 79.3 | +44 | [5] |

| 5M3H (S) | 5-methyl-3-heptanol | 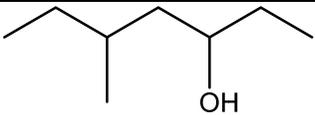 | 8 | 87.0 | +48 | [5] |
| --- | --- | --- | --- | --- | --- | --- |
| 5M2H (S) | 5-methyl-2-heptanol | 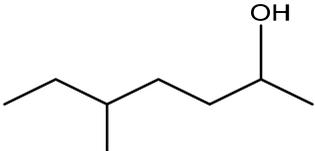 | 8 | 74.9 | +55 | [5] |
| 6M3H (S) | 6-methyl-3-heptanol | 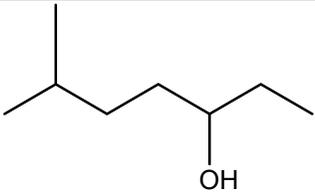 | 8 | 96.9 | +55 | [5] |
| 2E1H (P) | 2-ethyl-1-hexanol | 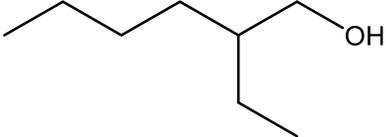 | 8 | 81.6 | +62 | [1] |
| 3,7 D1O (P) | 3,7-dimethyl-1-octanol | 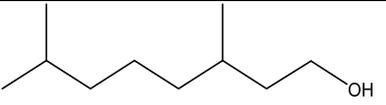 | 10 | 71.4 | +80 | [3] |
| 2B1O (P) | 2-butyl-1-octanol | 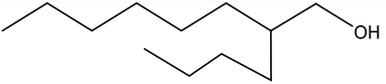 | 12 | 86.1 | +90 | [6] |
| 2H1D (P) | 2-hexyl-1-decanol | 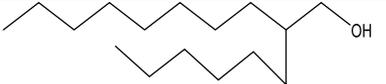 | 16 | 94.8 | +100 | [6] |

\* P: Primary alcohol; S: Secondary alcohol.
†Vertical ***b*** Shift factors used in **Figure 3** of the main context.

## 6. The "Living" polymer model and its extension to monohydroxy alcohols.

### 6.1 The "Living" polymer model for entangled polymers.

Cates [7] discussed how the dynamic reversibility of monomer unit affects the *entangled* polymer dynamics. Two major assumptions were made: (i) A chain can break with equal probability per unit time per unit length at all points in the chemical sequence of the polymer; and (ii) Two chains can combine with a rate proportional to the product of their concentrations. These two assumptions leads to a "living" polymer chain with an exponential distribution in chain length and a mean degree of polymerization of $\bar{N} = \sqrt{\frac{c_0 k_1}{2 k_2^2}}$, where $c_0$ is the molar concentration of the monomer, $k_1$ and $k_2$ are the reaction rate constant of monomer association and monomer dissociation. According to chemical reaction kinetics theory, the lifetime of the monomer in the "living" polymer is $\tau_0 = \frac{1}{k_2}$. The average chain breakage time of a "living" polymer with chain length $\bar{N}$ is $\tau_B = \frac{1}{k_2 \bar{N}}$ as a breakage of any one of the $\bar{N}$ bonds will lead to a breakup of the whole polymer chain.

Cates considered entangled polymers with $\bar{N} \gg N_e$, where $N_e$ is the entanglement degree of polymerization. The terminal relaxation of a polymer is $\tau_f$. If the $\bar{N}$ bonds are connected through covalent bonding, the end-to-end vector orientation dynamics can be described by reptation theory [8], $\tau_c$. In the limit of $\tau_B \geq \tau_c$, the reptation dynamics hold and the terminal relaxation time is controlled by $\tau_f$ with $\tau_f = \tau_c$.[7] Cates considered the cases of $\tau_B \ll \tau_c$, where the chain breakage occurs (multiple times) before a complete reptation. In this case, chain breakup and reformation facilitate the stress relaxation. As a result, the terminal relaxation time is much shorter than the reptation time, $\tau_f \ll \tau_c$. Cates argued that for *any* given segment, $x$, in the "living" polymer, the *fastest* path for stress relaxation of this segment (segment $x$) is through chain breakage of a sub-chain with a contour length is equal to the mean free path, $\lambda$, of this segment $x$ over a time period of $\tau_B$. As a result, the terminal relaxation or flow time of the "living" polymer is $\tau_f \approx \frac{1}{k_2 N_\lambda}$, where $N_\lambda$ is the number of Kuhn monomers of contour length $\lambda$. Note that prefactors are omitted in the scaling analysis.

According to the reptation model, the mean free path of any given segment in the chain is:

$$\lambda = \sqrt{D \tau_B} \qquad (S1)$$

with $D \sim \frac{k_B T}{\zeta \bar{N}}$ being the center-of-mass diffusion constant of the polymer and $\zeta$ the friction coefficient between Kuhn monomers. The contour length $\lambda$ corresponds to number of Kuhn monomers, $N_\lambda = \lambda / (\frac{b}{\sqrt{N_e}})$, where $b$ is the Kuhn length of the polymer and $N_e$ the entanglement number of Kuhn monomers. This gives the terminal relaxation time of the living polymer:

$$\tau_f \approx \frac{1}{k_2 \lambda / (\frac{b}{\sqrt{N_e}})} \sim \frac{\tau_B \bar{N} b}{\lambda \sqrt{N_e}} = \frac{\tau_B \bar{N} b}{\sqrt{D N_e \tau_B}} \sim (\tau_B \tau_c)^{1/2} \qquad (S2)$$

with $\tau_c = \frac{\langle L \rangle^2}{D} \sim \frac{b^2 \bar{N}^2}{N_e \frac{k_B T}{\zeta \bar{N}}} = \frac{\zeta b^2}{k_B T} \frac{\bar{N}^3}{N_e} = \tau_\alpha \frac{\bar{N}^3}{N_e}$ with $\tau_\alpha \equiv \frac{\zeta b^2}{k_B T}$ being the Kuhn segmental relaxation time (or the structural relaxation time), and $L = \frac{b \bar{N}}{\sqrt{N_e}}$ the contour length of the polymer. Eqn.S2 suggests the terminal relaxation time of the living polymer is a geometrical average of the chain breakage time, $\tau_B$, and its longest end-to-end vector relaxation time, $\tau_c$. Moreover, since chain breakage can happen randomly at any position of a long chain, the characteristic chain breakage time thus should follow a sharp Possion disitrubtion (central limit theorem), which gives a single-relaxation time relaxation process. We emphasize that near pure exponential decay of the terminal relaxation time of a "living" polymer has been observed and confirmed by a large amount of rheological measurements for worm-like micelles and other classical living polymeric systems [9,10].

### 6.2 Extension of the "living" polymer analysis to unentangled polymers

In unentangled polymers, the longest end-to-end vector relaxation is described by Rouse model with $\tau_c \sim \tau_\alpha \bar{N}^2$ [11]. In the fast breakup limit of $\tau_B \ll \tau_c$, the reorientational dynamics of a Rouse chain should also be affected by the chain breakup. The mean-free-path of a given monomer $x$, is $\lambda_R$, over a time scale of $\tau_B$:

$$\lambda_R \approx \sqrt{D_R \tau_B} \qquad (S3)$$

where $D_R = \frac{k_B T}{\zeta N_R}$ is the diffusion coefficient of a sub-chain of $\lambda_R$ in size. The number of of Kuhn monomers is $N_R \sim \lambda_R^2/b^2 \sim D_R \tau_B/b^2$. The stress relaxation through chain breakage is then:

$$\tau_f \approx \frac{1}{k_2 N_R} \approx \frac{1}{k_2 D_R \tau_B/b^2} \sim \frac{\bar{N} b^2}{D_R} \sim \frac{N_R}{\bar{N}} \tau_c \approx (\tau_B \tau_c)^{1/2} \qquad (S4)$$

which says the terminal time of a living unentangled chain is the geometrical average of the chain breakage time and the Rouse time of the polymer. Since $\tau_B \ll \tau_c$, this also leads to a speeding up in the terminal relaxation of the living polymer. The discussion in the main context is based on Eqn.S4.